\documentclass[aps]{revtex4-1}
\usepackage{graphicx}
\usepackage{amsmath}
\usepackage{amssymb}
\usepackage{tikz}
\usepackage{float}
\usepackage{sidecap}
\usepackage[caption = false]{subfig}
\usepackage{tabularx,booktabs}
  \newcolumntype{Y}{>{\centering\arraybackslash}X}
\usepackage[margin=1in]{geometry}
\usetikzlibrary{snakes}
\begin{document}

\title{Short-range interactions and
narrow resonances in effective field theory}
\author{Mohammad H. Alhakami}
 \affiliation{Nuclear Science Research Institute,
KACST, P.O. Box 6086, Riyadh 11442, Saudi Arabia}
 
\date{\today}
\begin{abstract}

We consider the effective field theory (EFT) treatment of two-body systems with narrow resonances. 
Within this approach, an $s$-wave scattering amplitude 
 can be expanded in powers of a typical momentum scale of a system $Q\ll \Lambda$, where $\Lambda$ represents a hard scale of a 
 scattering system and an energy difference $\delta \epsilon=|E-\epsilon_0|\ll \epsilon_0$, where $\epsilon_0$ is a resonance peak energy. 
It is shown that at leading order in the double expansion 
a universal form of a two-body scattering amplitude 
 is a sum of a Breit-Wigner term of order $Q^{-1}$,
a smooth background term of order $Q^0$, and an interference term 
of order $Q^0$. 
The techniques developed in this paper can be used to investigate the properties of narrow resonances that are 
produced by short-distance dynamics.

\end{abstract}

\pacs{Valid PACS appear here}
\maketitle

\section{INTRODUCTION}
The study of a two-body scattering problem with short-range interactions at low energies is of particular interest in 
nuclear and hadron physics. 
The amplitude of such systems is characterized by having a
\textit{universal form} in the sense that it is expressed in terms of a few physical parameters that do not depend on the details of 
short-range dynamics. Potential models used to derive the  scattering
amplitude are constructed based on the idea that  the details of the short-distance physics are unimportant 
as it cannot be resolved in the low-energy domain.  
This \textit{universal} 
result can also be 
reproduced in effective field theory (for reviews, see Refs. \cite{nrev,prev,erev}). 
This model-independent approach 
exploits the scale separation of a given system.
If one is interested in studying the low-energy dynamics in a physical system with soft scales $Q$ well separated from hard scales
$\Lambda$, an effective Lagrangian can be formulated as an expansion in powers of $Q$ with an infinite number of contact terms
(short-distance dynamics is hidden in the contact couplings) that are consistent with symmetries of the system.
However, a Lagrangian with an infinite number of terms is not predictive
and, as such, one has to implement power counting.
Power counting is an essential ingredient of any effective theory that is
necessary to organize terms in an effective Lagrangian according to their importance. 
For a low-energy two-body scattering, an effective Lagrangian is conveniently expressed 
in terms of scattering particles, as degrees
of freedom, interacting 
through a dimeron field that has quantum numbers of a resonance state \cite{dimeron}.
Here there are two limits that have a good separation of scales and thus one can build a powerful EFT. 
The first limit is (i) strong coupling between a dimeron field
and a scattering channel which in turn produces a virtual or bound state.
This is equivalent to
 the physics of summing bubble graphs to all orders \cite{ksw,ksw1,van,birse}. 
In this limit, the expansion parameters are the same. The second limit is (ii) weak coupling between a dimeron field
and a scattering channel
causes the production of a narrow resonance in the energy
spectrum. In this case, $\epsilon_0 \gg \Gamma$, where $\epsilon_0$ and $\Gamma$ are the energy and width of the resonance, respectively,
and this corresponds 
to a new low-energy scale. This in turn
provides a different expansion parameter from the strong-coupling limit \cite{ph,pd}.

For the above-mentioned strongly interacting systems, which generate bound/virtual states,
a scattering amplitude is expressed in terms of 
effective range expansion \cite{Bethe}. 
This result has also been reproduced in EFT; see, for example, Refs. \cite{ksw1,van}. 
In a similar manner, a scattering amplitude  
for low-energy two-body systems with short-range interactions that display narrow resonances 
can also be expressed in terms of 
a few physical scattering parameters and hence has a \textit{universal form}.  
Such systems have already been considered in Ref. \cite{bg} using EFT. It has been shown that, in the vicinity of the resonance region,
the \textit{universal form} of the amplitude at leading order in 
the double expansion of the theory ($Q$ and $\delta \epsilon$)
contains only
two distinct terms:
a Breit-Wigner term, which scales as $Q^{-1}$ and becomes significant at energies lying in the vicinity of the resonance energy $\epsilon_0$, 
and a smooth background term that scales as $Q^0$ and dominates for energies lying away from the resonance peak. 
It has been stated that interference terms, which involve both interactions, 
are suppressed by additional powers of $Q$ and $\delta \epsilon$.
However, as will be shown in this paper, the absence of such terms as leading contributions in the amplitude expansion is essentially a result 
of the improper use of EFT in Ref. \cite{bg}. 

This paper reconsiders the work of Gelman in \cite{bg} and is organized as follows. In Sec. II, we illustrate the scaling of the interaction parameters in powers of low-energy scales.
We present the form of the amplitude expansion 
of a two-body scattering system for energies lying in the vicinity of the resonance energy peak.
In Sec. III, the Lagrangian we use to reproduce the expansion of the amplitude in EFT is presented. 
We explicitly illustrate how to use a power counting scheme consistently to compute scattering amplitude in EFT. 
We show how to relate the couplings of the 
effective theory to the interaction parameters. The conclusion is given in Sec. IV.

\section{AN EXPANSION OF SCATTERING AMPLITUDE} 
Before discussing an EFT treatment for a low-energy two-body scattering system with a narrow resonance, 
let us start by recalling the relevant $s$-wave scattering amplitude for two nonrelativistic particles with spin-$0$ and common mass $M$,
normalized by a constant $\frac{4\pi}{M}$ \cite{bg},
\begin{equation}\label{bg1}
 \mathcal{A}=\frac{4\pi}{M}\left(\frac{e^{2i\delta^{(b)}_0}-1}{2ik}-\frac{1}{k}\frac{\Gamma/2}{E-\epsilon_0+i\Gamma/2}e^{2i\delta^{(b)}_0}\right),
\end{equation} 
where $k$ and $E$ are the relative momentum and total energy of the scattering particles in the center-of-mass frame, respectively.
The parameters $\epsilon_0$ and $\Gamma$ represent the energy and width of the resonance, respectively.  
The quantity $\delta^{(b)}_0$ defines the phase shift that results from background scattering.

The first term in Eq.~\eqref{bg1} gives rise to a background part of the scattering amplitude.
The second term, however, does not represent a pure Briet-Wigner amplitude as it contains an exponential factor $e^{2i\delta^{(b)}_0}$.
For energies lying in the vicinity of the resonance peak, i.e., $E\sim \epsilon_0$, the Briet-Wigner part is dominant
and the effect of background scattering becomes small; i.e., $\delta^{(b)}_0$ changes slowly. 
In this limit, the amplitude $\mathcal{A}$ given in Eq.~\eqref{bg1} can be expanded in powers of $\delta^{(b)}_0$ and $\delta \epsilon=|E-\epsilon_0|$.
To define the appropriate expansion of $\mathcal{A}$
for this system, let us first expand the exponential 
factor as
\begin{equation}\label{fb}
e^{2i\delta^{(b)}_0}\simeq 1+2i\delta^{(b)}_0-2\delta^{2(b)}_0+O\left(\delta^{3(b)}_0\right)=1-2ia_{bg}k-2a^2_{bg}k^2+O(a_{bg}^3k^3),
 \end{equation}
where $a_{bg}$ is the $s$-wave background scattering length, defined as  $$a_{bg}=-\lim_{k\to 0}  \frac{\mathrm{tan}\delta^{(b)}_0}{k}.$$

By substituting Eq.~\eqref{fb} into Eq.~\eqref{bg1}, one gets an expansion for 
\begin{equation}\label{bg2-0}
 \mathcal{A}=-\frac{4\pi}{M}\left(\frac{1}{k}\frac{\Gamma/2}{E-\epsilon_0+i\Gamma/2}+a_{bg} -2i a_{bg}\frac{\Gamma/2}{E-\epsilon_0+i\Gamma/2}-ia^2_{bg}k-2a^2_{bg}k\frac{\Gamma/2}{E-\epsilon_0+i\Gamma/2}+O(a_{bg}^3k^2)\right),
\end{equation}
in powers of $a_{bg}$, $k$, and $(\delta \epsilon)^{-1}$.
The expansion in Eq.~\eqref{bg2-0} is only valid for energies lying in the vicinity of the resonance energy, i.e., $E\sim \epsilon_0$.

To estimate the relative size of each term in the expansion of the amplitude $\mathcal{A}$ in Eq.~\eqref{bg2-0}, let us first
illustrate the scaling of the interaction parameters in powers of the low momentum of system $k$ that are generically  denoted  by $Q$:
the small background scattering length goes as $a_{bg}\sim Q^0$ and the energies $E$, $\epsilon_0$ scale as $Q^2$. In the vicinity of the low-lying resonance energy peak, $\delta \epsilon$ is taken as $Q^3$. 
For a narrow resonance $\Gamma/ \epsilon_0\ll 1$, the resonance width goes as $Q^3$.
It is worth mentioning that our scaling of the interaction parameters is similar to the
one presented in Ref. \cite{bg}. 

Based on the above scaling rules, 
the first term in Eq.~\eqref{bg2-0}, which represents a Breit-Wigner amplitude, scales 
as $Q^{-1}$. The second and third terms, which represent pure background and interference terms, respectively, scale as $Q^0$. 
The last two terms, which represent corrections to pure background and interference parts, respectively, scale as $Q$. 
Thus, the \textit{universal form} of a two-body scattering amplitude at leading order in the double expansion 
is the sum of the first three terms in the expansion Eq.~\eqref{bg2-0}. 
The third term in the above expansion was neglected in the corresponding result in Ref. \cite{bg} 
due to inconsistent use of the power-counting scheme, as will be shown below.
Our task now is to reproduce the expansion of the amplitude given in Eq.~\eqref{bg2-0} in an EFT. 

\section{AN EFT TREATMENT}
The relevant effective Lagrangian at leading order for 
a system of two spin-zero particles is 
\cite{bg}
 \begin{align}\label{bg4}
 \mathcal{L}=&\Psi^{\dag}\left(i \partial_0+\frac{\nabla^2}{2 M}\right)\Psi+\Phi^{\dag} \left(i \partial_0+\frac{ \nabla^2}{4 M}-\epsilon_0\right)\Phi
 -C_0 \left(\Psi^\dag \,\Psi\right)^2-g \left(\Phi^\dag\, \Psi\, \Psi+\Phi\, \Psi^\dag\, \Psi^\dag\right),
\end{align}
where $\Psi$ and $\Psi^{\dag}$  are the annihilation and creation operators for scattering particles of mass $M$.
The auxiliary operator $\Phi^{\dag}$ ($\Phi$) creates (destroys)  a dimeron field. The resonance parameter $\epsilon_0$ defines the residual energy of the dimeron.
The four-point particle-particle coupling constant $C_0$ measures the strength of background scattering.
The Yukawa-like coupling $g$ measures the coupling strength  between the dimeron (resonance) and the two particles in the scattering channel.

Here, producing a narrow resonance with a sharp peak is of interest.
Thus, we must treat $g$ as a small coupling; for a strongly coupled dimeron, only the bound or virtual states emerge
and there is no resonance.
The peak of the resonance becomes sharper as $g$ gets smaller.
If the coupling $C_0$ 
is weak, i.e., of natural size, then there will be no virtual or bound states.
In this case, the effect of the $s$-wave transition appears as a smooth background underneath the resonance peak.
In terms of the low-energy scale of the system $Q$, these couplings scale as 
\begin{align}\label{1}
 C_0\sim Q^0,~~~~g\sim Q.
\end{align}
For low energies, the kinetic energy of the two particles in the scattering channel, $E=k^2/M$,
and the resonance energy, $\epsilon_0$, are taken to be of order $Q^2$.
For energies lying in the vicinity of the resonance peak,
we take 
\begin{equation}\label{2}
\delta \epsilon=|E-\epsilon_0| \sim Q^3. 
\end{equation}
For this scattering problem, the theory has a double expansion in $\delta \epsilon/\epsilon_0$ and $k/\Lambda$, where $\Lambda$
defines the hard scale of the system.  
It should be stated that the power-counting scheme  introduced above 
is similar to the one employed 
in Ref. \cite{bg}. 

For coefficients of natural size,  
one can use minimal subtraction scheme (MS) to renormalize the theory \cite{ksw1}. 
In this scheme, the loop integral $I_0$ is defined as 
\begin{equation}\label{-3}
 I_0=\int \frac{{d^3}p}{(2\pi)^3}\frac{M}{k^2-p^2+i \epsilon}= -i\frac{M}{4 \pi} k\sim Q,
\end{equation}
from which it can be seen that there are no divergences. In this scheme, the factor $p$, which represents the internal momentum, inside the loop 
is converted to $k$, which represents the external momentum. 

Now our goal is to compute the scattering amplitude in EFT and equate the result to the expansion in Eq.~\eqref{bg2-0},
thereby fixing the couplings $C_0$ and $g$. 
According to the power
counting
scheme, 
one can expand the amplitude 
in EFT as 
\begin{equation}\label{AAA}
\mathcal{A}=\mathcal{A}_{-1}+\mathcal{A}_0+\mathcal{A}_1, 
\end{equation}
where terms in the expansion scale as $Q^{-1}$, $Q^0$, and $Q$, respectively, as implied by the subscript. 
To employ a power counting scheme consistently,
the Feynman graphs must be organized in a way that allows one to sum all graphs of a particular order to get the corresponding amplitude in the expansion.  
As illustrated below, the graphs with a particular set of couplings at the vertices will only contribute at a particular order in the expansion. 

Let us begin with the first term in the expansion Eq.~\eqref{AAA}. As shown in Fig.~\ref{dimeron}, this amplitude receives 
contributions from Feynman graphs that contain only the dimeron coupling at the vertices.
The sum of these diagrams,   
\begin{align}\label{a-}
 i\mathcal{A}_{-1}=-i\frac{g^2}{E-\epsilon_0}\left(1+I_0\frac{g^2}{E-\epsilon_0}+I_0\frac{g^2}{E-\epsilon_0}I_0\frac{g^2}{E-\epsilon_0}+...\right),
\end{align}
represents the full dimeron propagator.
 \begin{figure}[h!]
\begin{center}
\begin{tikzpicture}[scale=0.4]

\node at (-11,0) {$i\mathcal{A}_{-1}~=~$};

\draw[-,thick,line width=1pt](-8.3,0)--(-8.8,0.5);
\draw[<-,thick,line width=1pt](-8.8,0.5)--(-9.3,1);

\draw[-,thick,line width=1pt](-8.3,0)--(-8.8,-0.5);
\draw[<-,thick,line width=1pt](-8.8,-0.5)--(-9.3,-1);

\draw[-,thick,line width=1pt](-6.3,0)--(-5.8,0.5);
\draw[<-,thick,line width=1pt](-5.8,0.5)--(-5.3,1);

\draw[-,thick,line width=1pt](-6.3,0)--(-5.8,-0.5);
\draw[<-,thick,line width=1pt](-5.8,-0.5)--(-5.3,-1);

\draw[-,double,thick,line width=1pt](-8.3,0)--(-6.3,0) (-5.2,0) node[right]{$=$};

\draw[-,thick,line width=1pt](-3.3,0)--(-3.8,0.5);
\draw[<-,thick,line width=1pt](-3.8,0.5)--(-4.3,1);

\draw[-,thick,line width=1pt](-3.3,0)--(-3.8,-0.5);
\draw[<-,thick,line width=1pt](-3.8,-0.5)--(-4.3,-1);

\draw[-,thick,line width=1pt](-1.3,0)--(-0.8,0.5);
\draw[<-,thick,line width=1pt](-0.8,0.5)--(-0.3,1);

\draw[-,thick,line width=1pt](-1.3,0)--(-0.8,-0.5);
\draw[<-,thick,line width=1pt](-0.8,-0.5)--(-0.3,-1);

\draw[-,thick,line width=1pt](-3.3,0)--(-1.3,0) (-0.2,0) node[right]{$+$};

\draw[-,thick,line width=1pt](1.7,0)--(1.2,0.5);
\draw[<-,thick,line width=1pt](1.2,0.5)--(0.7,1);

\draw[-,thick,line width=1pt](1.7,0)--(1.2,-0.5);
\draw[<-,thick,line width=1pt](1.2,-0.5)--(0.7,-1);

\draw[thick,line width=1pt] (3.7,0) arc (180:360: 1cm) (4.2,1) node[right]{$<$} ;
\draw [thick,line width=1pt] (3.7,0) arc (180:0: 1cm) (4.2,-1) node[right]{$>$};

\draw[-,thick,line width=1pt](1.7,0)--(3.7,0) (4.8,0);

\draw[-,thick,line width=1pt](5.7,0)--(7.7,0) (8.8,0) node[thick,right]{$~~+$};

\draw[-,thick,line width=1pt](7.7,0)--(8.2,0.5);
\draw[<-,thick,line width=1pt](8.2,0.5)--(8.7,1);

\draw[-,thick,line width=1pt](7.7,0)--(8.2,-0.5);
\draw[<-,thick,line width=1pt](8.2,-0.5)--(8.7,-1) ;


\draw[-,thick,line width=1pt](11.7,0)--(11.2,0.5);
\draw[<-,thick,line width=1pt](11.2,0.5)--(10.7,1);

\draw[-,thick,line width=1pt](11.7,0)--(11.2,-0.5);
\draw[<-,thick,line width=1pt](11.2,-0.5)--(10.7,-1);

\draw[-,thick,line width=1pt](11.7,0)--(13.7,0);

\draw[thick,line width=1pt] (13.7,0) arc (180:360: 1cm) (14.2,1) node[right]{$<$} ;
\draw [thick,line width=1pt] (13.7,0) arc (180:0: 1cm) (14.2,-1) node[right]{$>$};

\draw[-,thick,line width=1pt](15.7,0)--(17.7,0);

\draw[thick,line width=1pt] (17.7,0) arc (180:360: 1cm) (18.2,1) node[right]{$<$} ;
\draw [thick,line width=1pt] (17.7,0) arc (180:0: 1cm) (18.2,-1) node[right]{$>$};

\draw[-,thick,line width=1pt](19.7,0)--(21.7,0)  (22.8,0)node[right]{$+...$};

\draw[-,thick,line width=1pt](21.7,0)--(22.2,0.5);
\draw[<-,thick,line width=1pt](22.2,0.5)--(22.7,1);

\draw[-,thick,line width=1pt](21.7,0)--(22.2,-0.5);
\draw[<-,thick,line width=1pt](22.2,-0.5)--(22.7,-1);

\end{tikzpicture}
\caption{The Feynman graphs that contribute to $\mathcal{A}_{-1}$. This amplitude, which represents
the full dimeron propagator (double solid lines), gets a contribution from the bare dimeron propagator (solid line) dressed with particle bubbles.
The scattering particles are represented by solid lines with arrows.}
\label{dimeron}
\end{center}
\end{figure}
According to the above power counting rules, terms in Eq.~\eqref{a-} are of the same order, $Q^{-1}$, as the first term in the expansion. As a
consequence, one must sum up all iterated diagrams to get \cite{ph,pd,bg}

\begin{align}\label{b-}\nonumber
\mathcal{A}_{-1}&=-\frac{g^2}{E-\epsilon_0-g^2 I_0 }\\ 
&=-\frac{g^2}{E-\epsilon_0+i(M/4\pi) g^2 k }.
\end{align}
The dimeron coupling $g^2$ can now be fixed by equating Eq.~\eqref{b-} with the first term in the expansion Eq.~\eqref{bg2-0}. This yields 
\begin{equation}\label{fixg}
g^2=\frac{4\pi}{M}\frac{1}{k}\Gamma(E)/2. 
\end{equation}
The amplitude in Eq.~\eqref{b-}, 
which can be expressed 
in terms of the resonance decay width $\Gamma$ as
\begin{align}\label{A-1}
\mathcal{A}_{-1}=-\frac{4\pi}{M}\frac{1}{k}\frac{\Gamma/2}{E-\epsilon_0+i\Gamma/2 },
\end{align}
represents the Breit-Wigner resonance
and is counted as $Q^{-1}$. 
By introducing the dimensionless coupling  
$g^{\prime}=g\sqrt{M/4\pi}$,  
the energy-dependent decay width of the resonance is
\begin{equation}\label{decayF}
 \Gamma(E)=2 g^{ \prime 2}k=2g^{\prime 2}\sqrt{ME}.
\end{equation}
At the threshold, $\Gamma(E)$ is very small, while
at energies well above
the threshold, the number of states will increase and $\Gamma(E)$ becomes large.
At $E=\epsilon_0$, this gives the physical width of the resonance   $\Gamma(\epsilon_0)$. In our power-counting
scheme, the size of $\Gamma(E)$
is $Q^3$ and this indicates that the resonance is very narrow.
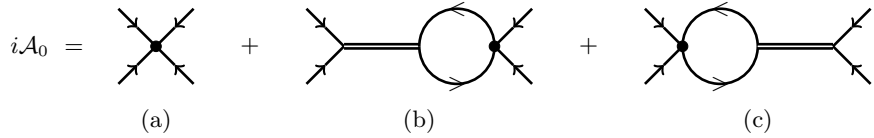
\begin{figure}[h!]
\begin{center}
\begin{tikzpicture}[domain=1:5,scale=0.5]
\node at (-3,0) {$i\mathcal{A}_0~=~$};

\draw[-,thick,line width=1pt](-.3,0)--(-.8,0.5);
\draw[<-,thick,line width=1pt](-.8,0.5)--(-1.3,1);

\draw[-,thick,line width=1pt](-.3,0)--(-.8,-0.5);
\draw[<-,thick,line width=1pt](-.8,-0.5)--(-1.3,-1);
\draw[fill] (-0.3,0) circle (0.14);
\node at (-.3,-2) {(a)};

\draw[-,thick,line width=1pt](-.3,0)--(.2,0.5);
\draw[<-,thick,line width=1pt](.2,0.5)--(.7,1);

\draw[-,thick,line width=1pt](-.3,0)--(.2,-0.5);
\draw[<-,thick,line width=1pt](.2,-0.5)--(.7,-1);

\node at (2.2,0) {$+$};

\draw[-,thick,line width=1pt](4.7,0)--(4.2,0.5);
\draw[<-,thick,line width=1pt](4.2,0.5)--(3.7,1);

\draw[-,thick,line width=1pt](4.7,0)--(4.2,-0.5);
\draw[<-,thick,line width=1pt](4.2,-0.5)--(3.7,-1);
\draw[-,double,thick,line width=1pt](4.7,0)--(6.7,0);

\node at (6.7,-2) {(b)};

\draw[thick,line width=1pt] (6.7,0) arc (180:360: 1cm) (7.2,1) node[right]{$<$} ;
\draw [thick,line width=1pt] (6.7,0) arc (180:0: 1cm) (7.2,-1) node[right]{$>$};

\draw[-,thick,line width=1pt](8.7,0)--(9.2,0.5);
\draw[<-,thick,line width=1pt](9.2,0.5)--(9.7,1);

\draw[-,thick,line width=1pt](8.7,0)--(9.2,-0.5);
\draw[<-,thick,line width=1pt](9.2,-0.5)--(9.7,-1) ;
\draw[fill] (8.7,0) circle (0.14);

\draw[-,thick,line width=1pt](13.7,0)--(13.2,0.5);
\draw[<-,thick,line width=1pt](13.2,0.5)--(12.7,1);

\draw[-,thick,line width=1pt](13.7,0)--(13.2,-0.5);
\draw[<-,thick,line width=1pt](13.2,-0.5)--(12.7,-1);
\draw[fill] (13.7,0) circle (0.14);

\node at (11.2,0) {$+$};

\draw[thick,line width=1pt] (13.7,0) arc (180:360: 1cm) (14.2,1) node[right]{$<$} ;
\draw [thick,line width=1pt] (13.7,0) arc (180:0: 1cm) (14.2,-1) node[right]{$>$};
\draw[-,double,thick,line width=1pt](15.7,0)--(17.7,0);

\node at (15.7,-2) {(c)};

\draw[-,thick,line width=1pt](17.7,0)--(18.2,0.5);
\draw[<-,thick,line width=1pt](18.2,0.5)--(18.7,1);

\draw[-,thick,line width=1pt](17.7,0)--(18.2,-0.5);
\draw[<-,thick,line width=1pt](18.2,-0.5)--(18.7,-1);

\end{tikzpicture}
\caption{The amplitude $A_0$ arises from (a) a tree graph with $C_0$ at the vertex and (b)-(c)  
two one-loop graphs that consist of the full dimeron propagator and have one $C_0$ at the vertices.
The notation is the same as in Fig.~\ref{dimeron}.}
\label{contact}
\end{center}
\end{figure}
While the amplitude $\mathcal{A}_{-1}$ cannot be computed perturbatively in EFT [see Eq.~\eqref{b-} and the relevant Feynman graphs shown in 
Fig.~\ref{dimeron}], the amplitudes $\mathcal{A}_0$ and $\mathcal{A}_1$ can be, as shown in Figs.~\ref{contact}--\ref{correct}. 
The amplitude $\mathcal{A}_0$ is obtained by adding the contribution from 
the tree diagram with $C_0$ at the vertex [Fig.~\ref{contact}(a)] with the contribution coming from the two one-loop graphs.
These two one-loop graphs include  
the full dimeron propagator and have one $C_0$ at the vertices [Figs.~\ref{contact}(b) and \ref{contact}(c)]. This gives
\begin{align}\label{0a1}
 i\mathcal{A}_0= -i \left(C_0+2 C_0 I_0 \frac{g^2}{E-\epsilon_0-g^2I_0 }\right),
\end{align}
where the first term represents the background scattering and the second term represents the mixing between Breit-Wigner and background
scattering. Both contributions scale as $Q^0$. 
Using Eqs.~\eqref{-3} and \eqref{fixg},  Eq.~\eqref{0a1} becomes
\begin{align}\label{0a}
\mathcal{A}_0= - \left(C_0- 2i C_0\frac{\Gamma/2}{E-\epsilon_0+i\Gamma/2 }\right).
\end{align}
One can fix the four-point coupling $C_0$ by equating Eq.~\eqref{0a} to the second and third terms in the expansion Eq.~\eqref{bg2-0}.
This gives
\begin{equation}\label{fixc}
C_0=\frac{4\pi}{M}a_{bg}. 
\end{equation}
In terms of the background scattering length, Eq.~\eqref{0a} can be written as
\begin{align}\label{00a}
\mathcal{A}_0= - \frac{4\pi}{M}\left(a_{bg}- 2i a_{bg}\frac{\Gamma/2}{E-\epsilon_0+i\Gamma/2 }\right).
\end{align}

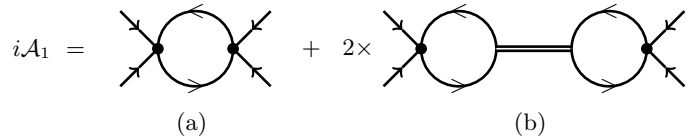
\begin{figure}[h!]
\begin{center}
\begin{tikzpicture}[scale=0.5]
\node at (4,0) {$i\mathcal{A}_1~=~$};

\draw[-,thick,line width=1pt](6.7,0)--(6.2,0.5);
\draw[<-,thick,line width=1pt](6.2,0.5)--(5.7,1);

\draw[-,thick,line width=1pt](6.7,0)--(6.2,-0.5);
\draw[<-,thick,line width=1pt](6.2,-0.5)--(5.7,-1);
\draw[fill] (6.7,0) circle (0.14);

\draw[thick,line width=1pt] (6.7,0) arc (180:360: 1cm) (7.2,1) node[right]{$<$} ;
\draw [thick,line width=1pt] (6.7,0) arc (180:0: 1cm) (7.2,-1) node[right]{$>$};

\node at (7.6,-2) {(a)};

\draw[-,thick,line width=1pt](8.7,0)--(9.2,0.5);
\draw[<-,thick,line width=1pt](9.2,0.5)--(9.7,1);

\draw[-,thick,line width=1pt](8.7,0)--(9.2,-0.5);
\draw[<-,thick,line width=1pt](9.2,-0.5)--(9.7,-1) ;
\draw[fill] (8.7,0) circle (0.14);

\node at (11.2,0) {$~~+~~2\times$};
\draw[-,thick,line width=1pt](13.7,0)--(13.2,0.5);
\draw[<-,thick,line width=1pt](13.2,0.5)--(12.7,1);

\draw[-,thick,line width=1pt](13.7,0)--(13.2,-0.5);
\draw[<-,thick,line width=1pt](13.2,-0.5)--(12.7,-1);
\draw[fill] (13.7,0) circle (0.14);

\draw[thick,line width=1pt] (13.7,0) arc (180:360: 1cm) (14.2,1) node[right]{$<$} ;
\draw [thick,line width=1pt] (13.7,0) arc (180:0: 1cm) (14.2,-1) node[right]{$>$};
\draw[-,double,thick,line width=1pt](15.7,0)--(17.7,0);

\node at (16.6,-2) {(b)};

\draw[thick,line width=1pt] (17.7,0) arc (180:360: 1cm) (18.2,1) node[right]{$<$} ;
\draw [thick,line width=1pt] (17.7,0) arc (180:0: 1cm) (18.2,-1) node[right]{$>$};

\draw[-,thick,line width=1pt](19.7,0)--(20.2,0.5);
\draw[<-,thick,line width=1pt](20.2,0.5)--(20.7,1);

\draw[-,thick,line width=1pt](19.7,0)--(20.2,-0.5);
\draw[<-,thick,line width=1pt](20.2,-0.5)--(20.7,-1);
\draw[fill] (19.7,0) circle (0.14);
\end{tikzpicture}
\caption{The amplitude $\mathcal{A}_1$ receives contributions from all Feynman graphs with two $C_0$'s at the vertices.
In this figure, (a) provides a correction to Fig.~\ref{contact}(a) and (b) provides a correction to Figs.~\ref{contact}(b) and \ref{contact}(c).
The notation is the same as in Fig.~\ref{dimeron}.}\label{correct}
\end{center}
\end{figure}

The amplitude $\mathcal{A}_1$, which provides corrections to $\mathcal{A}_0$, arises from a one-loop graph with two $C_0$'s at the vertices 
in addition to two two-loop graphs that consist of the full dimeron propagator and two $C_0$'s at the vertices [see Figs.~\ref{correct}(a) and \ref{correct}(b)]:
\begin{align}\label{1a}
i\mathcal{A}_1= -i \left(C^2_0I_0+2C^2_0I^2_0\frac{g^2}{E-\epsilon_0-g^2I_0}\right),
\end{align}
where the first and second terms are corrections to the background and interference terms in $\mathcal{A}_0$, respectively.
Using Eqs.~\eqref{-3}, \eqref{fixg}, and \eqref{fixc}, Eq.~\eqref{1a} becomes
\begin{align}\label{2a}
\mathcal{A}_1= \frac{4\pi}{M}\left(ia^2_{bg}k+ 2a^2_{bg}k\frac{\Gamma/2}{E-\epsilon_0+i\Gamma/2 }\right).
\end{align}

By substituting Eqs.~\eqref{A-1}, \eqref{00a}, and \eqref{2a}
into Eq.~\eqref{AAA}, one gets 
\begin{align}\label{AAAA}
 \mathcal{A}
 =-\frac{4\pi}{M}\left(\frac{1}{k}\frac{\Gamma/2}{E-\epsilon_0+i\Gamma/2}+a_{bg} -2i a_{bg}\frac{\Gamma/2}{E-\epsilon_0+i\Gamma/2}-ia^2_{bg}k-2a^2_{bg}k\frac{\Gamma/2}{E-\epsilon_0+i\Gamma/2}\right),
 \end{align}
which has the exact form as the expansion of the amplitude in Eq.~\eqref{bg2-0}. 
The first three terms in Eq.~\eqref{AAAA} [or Eq.~\eqref{bg2-0}], which represent the Breit-Wigner resonance, background scattering, and their interference,
appear as leading contributions in the double expansion in powers of $Q$ and $\delta \epsilon$. 
The power-counting scheme implemented here leads to an enhancement of the interference term that have been neglected in Ref. \cite{bg}. 
The corresponding result presented in \cite{bg}  was obtained by adding only the contributions 
from the Feynman diagrams given in Figs.~\ref{dimeron} and \ref{contact}(a).
The other Feynman graphs shown in Figs.~\ref{contact}(b) and \ref{contact}(c) were neglected
in \cite{bg}. These Feynman graphs represent the interference contributions 
and scale as $Q^0$ like the background contribution, which indicates the inconsistent use of the power counting scheme in Ref. \cite{bg}.

\section{CONCLUSION}
Scattering systems with short-range interactions and narrow resonances have been discussed 
in this paper. In Sec. II, the scaling of the interaction parameters in terms of the low-energy scale $Q$ has been explicitly shown. 
For energies lying 
in the vicinity of the resonance energy $E\sim \epsilon_0$,
the expansion of the two-body $s$-wave  scattering amplitude is presented. 
It has been shown that at leading order in powers of $Q$ and $\delta \epsilon$,
the amplitude is a sum of a Breit-Wigner term of order $Q^{-1}$,
a smooth background term of order $Q^0$, and an interference term 
of order $Q^0$, where $Q$ is a small momentum scale; see  Eq.~\eqref{bg2-0}.
An EFT treatment for these scattering systems has also been considered in this paper. 
In Sec. III, an effective Lagrangian is constructed as a double expansion in powers of
$Q$ and $\delta \epsilon$. At leading order, the Lagrangian consists of two contact couplings: a four-point coupling $C_0$, which measures the strength 
of the background scattering, and a three-point coupling $g$, which measures the strength of interactions between the dimeron field and the scattering channel.
In our work, $g$ was taken as $Q$ to produce a narrow resonance with a sharp peak and
$C_0$ was treated as $Q^0$ to make the effect of short-range interaction appear as a smooth background underneath the resonance peak.
We have shown how to use a power counting scheme consistently 
in computing scattering amplitude in EFT. The couplings $C_0$ and $g$ are related to the interaction parameters $a_{bg}$ and $\Gamma$
by comparing the EFT result with the expansion of the amplitude given in Eq.~\eqref{bg2-0}.
Our result can be used to investigate the properties of very near-threshold states such as the $X(3872)$ \cite{bel,br1,br2,nev,han1,nora, han2,rep, alhakami}.

\begin{center}
\textbf{ACKNOWLEDGMENTS}
\end{center}
I would like to thank Michael C. Birse for helpful discussions and encouragement.

\end{document}